# THE EFFECT OF AQUEOUS ALTERATION AND METAMORPHISM IN THE SURVIVAL OF PRESOLAR SILICATE GRAINS IN CHONDRITES

**Josep Mª Trigo-Rodriguez**

Institut of Space Sciences (CSIC), Campus UAB, Facultat de Ciències,

Torre C5-parell-2ª, 08193 Bellaterra, Barcelona, Spain. E-mail: trigo@ieec.uab.es

Institut d'Estudis Espacials de Catalunya (IEEC), Edif.. Nexus,

c/Gran Capità, 2-4, 08034 Barcelona, Spain

and

**Jürgen Blum**

Institut fürGeophysik und extraterrestrische Physik, Technische Universität Braunschweig,

Mendelssohnstr. 3, 38106 Braunschweig, Germany. E-mail: j.blum@tu-bs.de

## ABSTRACT

**Relatively small amounts (typically between 2-200 parts per million) of presolar grains have been preserved in the matrices of chondritic meteorites. The measured abundances of the different types of grains are highly variable from one chondrite to another, but are higher in unequilibrated chondrites that have experienced little or no aqueous alteration and/or metamorphic heating than in processed meteorites. A general overview of the abundances measured in presolar grains (particularly the recently identified presolar silicates) contained in primitive chondrites is presented. Here we will focus on the most primitive chondrite groups, as typically the highest measured abundances of presolar grains occur in primitive chondrites that have experienced little thermal metamorphism. Looking at the most aqueously altered chondrite groups, we find a clear pattern of decreasing abundance of presolar silicate grains with increasing level of aqueous alteration. We conclude that the measured abundances of presolar grains in altered chondrites are strongly biased by their peculiar histories. Scales quantifying the intensity of aqueous alteration and shock metamorphism in chondrites could correlate with the content in presolar silicates. To do this it would be required to infer the degree of destruction or homogenization of presolar grains in the matrices of primitive meteorites. To get an unbiased picture of the relative abundance of presolar grains in the different regions of the protoplanetary disk where first meteorites consolidated, future dedicated studies of primitive meteorites, IDPs, and collected materials from sample-return missions (like e.g. the planned Marco Polo) are urgently required.**

# INTRODUCTION: PRESOLAR GRAIN ABUNDANCES

An evaluation of the primordial abundance of presolar grains in the materials mixed in the protoplanetary disk at the time the chondritic meteorites formed deserves all our efforts because it allows a comparison with the abundances expected from theoretical models. Such models usually take into account the abundances expected from the galactic chemical evolution and the stellar dust production in the solar environment where the Solar System formed. It is obvious that for gaining insight into these processes we should look at the most primitive meteorites that have arrived on Earth. These are among the oldest rocks formed in our planetary system, and represent samples derived from undifferentiated parent bodies that have not experienced significant heating after accretion. This condition was probably achieved by small size asteroids where radiogenic heating was released at a rate that the temperature of the materials never reached the melting limit (see e.g. Trieloff et al., 2003; Hutchison et al., 2006). These rocks preserved the components that were forming the protoplanetary disk before the accretion of planets, but also suffered different secondary alteration processes in their parent bodies, mostly aqueous alteration and thermal metamorphism.

Up to date 14 chondrite groups have been identified (see e.g. Weisber et al., 2006) designated with a one or two letter symbol and have a characteristic chemical composition. The evident chemical differences among the chondrite groups led to the idea that each group represents rocks coming from a different reservoir (see eg. Hutchison, 2004). This idea has been recently reinforced because relatively few chondrite breccias exist that contain clasts belonging to different chondrite groups (Bischoff et al, 2006). Primitive asteroids are covered by rubble produced by continuous impacts that have excavated and fragmented their surfaces. Consequently, chondrite breccias are mainly formed by the compaction of the asteroidal regolith under the action of impacts, whereas at the typical encounter velocities the original projectile material would be preferentially vaporized.

Inside each chemical group there are meteorites that belong basically to 6 different petrologic types. Types 1 and 2 are restricted to those groups affected by aqueous alteration at low temperatures. It is believed that these groups come from small asteroids that never suffered significant radioactive heating. Type 3 is represented by meteorites that are in the intersection of two processes: aqueous alteration and thermal metamorphism (see e.g. Hutchison, 2004). In general type 3 chondrites experience little aqueous alteration and metamorphism, so they are considered the most pristine chondrites available in meteorite collections. Thermal metamorphism increasingly affected petrologic types 3 to 6, type 6 being the most affected by in-situ parent body heating. A comprehensive review on the thermal metamorphism experienced by all chondrite groups has been recently published (Huss et al., 2006).

We should first remark that the degree of thermal metamorphism and aqueous alteration is highly variable in the different chondrite groups. For example, the CI and CM groups suffered extensive aqueous alteration, but for the most part escaped thermal metamorphism (only a few CMs evidence heating over several hundred Kelvin). In fact, the CI and CM chondrites are water-rich, and secondary minerals as consequence of the pervasive alteration of primary phases are common. On the other hand, CO, CV, and CR chondrites suffered much less severe aqueous alteration, but some CRs are moderately aqueously altered. Among these groups, CO and CV are particularly good candidates to find presolar grains as they experienced moderately small heating. Thermal metamorphic grades for both groups are ranging from low (3.0) to nearly type 4 (Russell et al., 1998; Bonal et al., 2006, 2007). Chondrites that we would consider "pristine" are quite rare. One would be the CO chondrite ALHA77307, another is the CM-like ungrouped Acfer094 (Rubin et al., 2007). To find pristine chondrites among the different groups of ordinary chondrites (H, L, and LL) is also complicated, but some examples exist like e.g. LL3.0 Semarkona, and LL3.1 Bishunpur (Mostefaoui et al., 2004). Most of the meteorites integrating these chondrite groups can show that aqueous alteration and thermal metamorphism change the abundances of presolar silicates. In any case, we should remark that the initial stellar grain abundances in the fine-grained matrix of each group was probably not the same (Huss et al., 2003).

Huss (1990) performed a study of the abundance of interstellar diamond and silicon carbide (SiC) in chondrites reaching the conclusion that the metamorphic destruction is responsible for the absence of these grains in most chondrites. Huss and Lewis (1995) confirmed such scenario, but after controlling for the affects of thermal metamorphism by looking at the most primitive members of each group, they suggested the existence of some pre-metamorphic variation from class to class. They inferred that this variation might reflect thermal processing in the protoplanetary disk. Huss et al. (1996) expanded on the interpretation of these data by including the carrier of the "normal planetary" noble gases. Huss et al. (2003) expanded the database to include additional carbonaceous chondrites. While chemical procedures allow the separation of diamonds, SiC and other chemically resistant grains contained in chondrites (see e.g. Ott, 2007), these are not the only presolar grains contained in chondrites. Recent use of the nanoSIMS for studying the polished thin sections of primitive chondrites allowed the in-situ identification of several hundred presolar silicates in an increasing number of meteorites (see e.g. Nguyen, 2005). These presolar silicates are usually identified by the large (sometimes huge) isotopic anomalies exhibited relative to the other components of the matrix that are surrounding them (Nguyen and Zinner, 2004; Mostefaoui and Hoppe, 2004; Nagashim et al., 2004).

THE ABUNDANCE OF PRESOLAR SILICATES IN PRIMITIVE CHONDRITE GROUPS

Presolar silicates are embedded into the fine-dust matrices of carbonaceous chondrites. These meteorites can contain up to 12% of water by weight, which, given the obvious rocky nature of these rocks, is bound in secondary minerals. These aqueously altered minerals are mainly clays with serpentine, carbonates, sulfates, sulfides, phosphates, and oxides. Most (if not all) of these secondary minerals formed from a water solution that was soaking the parent bodies of carbonaceous chondrites (for a detailed discussion see Jewitt et al., 2007). Mobilization of water through the fine dust materials that form the matrices of these chondrites produced the alteration of the original minerals. Particularly quick alteration occurs for those tiny mineral grains that are forming the matrix of carbonaceous chondrites (see e.g. Trigo-Rodríguez et al., 2006; Rubin et al., 2007).

Presolar silicates were incorporated in these rocks as tiny mineral grains with typical sizes smaller than one µm (Zinner, 2003; Zinner et al., 2003). Due to their small sizes they can be particularly affected by aqueous alteration, which tends to equilibrate the composition of the grains with the surrounding materials forming the matrices (Brearley, 1996). In fact, even though this equilibration was not complete, it renders the search of presolar silicates in aqueously altered chondrites very difficult. The reason is that current searches for presolar silicates using the nanoSIMS involve the identification of isotopic anomalies in these grains relative to the background matrix minerals. The typically studied isotopic ratios compared in nanoSIMS images are $^{17}O/^{16}O$, $^{18}O/^{16}O$, $^{28}Si/^{16}O$, $^{27}Al/^{16}O$ (see e.g. Zinner et al., 2003). It is important to remark that aqueous alteration followed by thermal metamorphism dilutes these isotopic anomalies into the surrounding matrices, complicating the identification of presolar silicates in aqueously altered samples.

We will discuss briefly the main characteristics of the different chondrite groups with identified pristine or moderately pristine meteorites, discussing the effect that collisional compaction and water could have had in affecting the preservation of presolar grains in their matrices. We start by discussing the CI group that is considered the prototype of starting material with chemical abundance similar to the Sun.

### The CI chondrite group

Chondrites belonging to this group are characterized by being water-rich, but are different from other chondrites in being chondrule-free. Because of the close match between their chemical compositions as that of the solar photosphere, CI chondrites are considered to be the group most representative of the bulk composition of the solar system (Anders and Grevesse, 1989). It suggests that this group of chondrites best preserves the bulk chemical composition of presolar material, but probably these meteorites suffered significant compaction and the extreme effects of aqueous alteration. In fact, the matrices of these meteorites are extremely fine-grained hydrous silicates like e.g. clays and oxides (see e.g. Brearley and Jones, 1998).

The primitiveness of this group is well exemplified by members such as Orgueil that have high abundances of presolar grains, particularly refractory ones: e.g. diamond, graphite and SiC (Huss, 1997). An identification of presolar silicates has been not obtained yet, but measuring their abundance in this meteorite would be a good test for our understanding of the role of aqueous alteration in the destruction of grain properties in the CCs matrices. Future accurate, and surface-representative studies of the abundances of isotopically anomalous presolar grains by NanoSIMS isotopic mapping are required. They can provide additional insights on the degree of preservation as a function of aqueous alteration. If aqueous alteration plays a role in destroying presolar silicates, then the abundances of olivine and pyroxene presolar grains should be minimum for this group.

### The CM group

CM chondrites are also aqueously altered rocks that contain ~9 wt.% water bound in phyllosilicates. Other alteration minerals probably formed by precipitation of an aqueous solution; e.g. the clumps of serpentine-tochilinite intergrowths, pentlandite and pyrrhotite (Rubin et al., 2007). Not all CM chondrites exhibit the same degree of aqueous alteration, and even the same meteorite can exhibit clasts with different alteration (e.g. Murchison or Murray).

Rubin et al. (2007) proposed a numerical alteration sequence for CM chondrites.

Hypothetically unaltered CM chondrites would be similar to the ungrouped Acfer094 (Trigo-Rodríguez et al., 2006). The chemical composition of Acfer 094 was compared with the mean for the CM group in Table 6 of Rubin et al. (2007) showing that this meteorite is CM-related. In fact, the calculated presolar silicate abundance in Acfer 094 (Table 1) is much higher than that of other presolar phases found in meteorites (Nguyen et al., 2005). Rubin et al. (2007) classified this meteorite as C3.0 in their new numerical alteration sequence. The pristine nature of Acfer094 meteorite is shown in Fig. 1. A fine-grained uncompacted, unprocessed, and extensive matrix, such as this exemplified in Fig. 1, seems to be favourable to obtain good results in searching presolar grains by ion beam rastering using NanoSIMS.

Another ungrouped meteorite related to the CM group is Tagish Lake, but it is representing a body poor in presolar silicates. We should note that Tagish Lake is significantly richer in C and volatile elements than the CM group (Grady et al., 2002). The parent body of Tagish Lake has been identified as a D-type dark asteroid formed in the outer part of the main belt (Hiroi et al., 2001). Table 1 shows that the measured presolar silicate abundance of Tagish Lake is very low, probably reflecting the extreme degree of aqueous alteration experienced by its parent body. In fact, this type of primitive asteroids could have experienced significant hydration as consequence of their primordial (nebular) volatile-rich nature.

<u>The CO group</u>

The members of the CO group are petrologic type 3, no aqueously altered representatives of types 1 or 2 have been identified so far. Nevertheless, water played a role in the alteration history of the CO chondrites. However, the full group is subclassified as a metamorphic sequence. Scott and Jones (1990) and Sears et al. (1991) completed the classification of this group, defining the main petrologic types. Those relatively pristine, highly unequilibrated COs are classified as subgroup CO 3.0, while those experiencing a higher degree of thermal metamorphism are classified as 3.8 (Chizmadia et al., 2002). In between this sequence we found different degrees of metamorphic products.

The role of water in the group was minor, and still remains debated. In fact, micrometer-sized silicate grains are well preserved in highly unequilibrated CO 3.0 members like ALHA77307 (Scott and Krot, 2003). Despite of the pristine nature of the matrix of ALHA77307 evidenced in Fig. 2, we should remark that this meteorite experienced some terrestrial weathering. However, this meteorite also contains phyllosilicates and anhydrite that were probably produced in the parent body, although Brearley (1993) proposed that they might be weathering products from Antarctica. Other authors, like e.g. Keller and Buseck (1990), pointed out that the matrix of Lancé, a witnessed fall of petrologic type 3.4, also contains phyllosilicates that should have necessarily formed in the parent body. This picture would be clear if the CO parent body had little amounts of water or was quickly released as consequence of thermal metamorphism. This peculiarity relative to other groups (e.g. CM or CI) would make the highly unequilibrated COs good targets for searching for presolar silicates. In fact, the abundances of other types of presolar grains (particularly diamond, SiC and graphite) were noticed to decrease systematically with increasing petrologic type (Huss, 1990; Huss et al., 2003).

### The CV group

Many members of the CV group have been extensively affected by secondary processes, including alkali-halogen Fe metasomatism, aqueous alteration, thermal metamorphism and shock metamorphism (Krot et al., 1995). Moreover, the CV3 chondrites may have lost most of the fragile presolar grains before they ever accreted due to high temperatures in the nebula (e.g., Huss et al., 2003, Huss, 2004). Bonal et al. (2004, 2006) performed Raman spectroscopy of organic matter to better determine the petrologic type of some CVs. The CV group is currently divided into three subgroups (two oxidized and one reduced) on the basis of the secondary alteration features. In particular the CV3 reduced subgroup is represented by Leoville, Efremovka, and Vigarano, among others. Consequently, searches for presolar grains in these previously mentioned meteorites have been performed because they are among the most unaltered meteorites of the CV group, but still show evidence of having experienced significant thermal metamorphism. This is currently explained by the fact that the reduced subgroup suffered little or almost no aqueous alteration, different from the two oxidized subgroups (Huss et al., 2006). The measured abundances of presolar grains also indicate that Leoville is less metamorphosed than the oxidized Mokoia, followed by Vigarano, Axtell, and Allende (Huss et al., 2003).

### The CR group

The members of this group also suffered different degrees of aqueous alteration. Renazzo and Al Rais are members of this group (named after the first one) although were originally identified as CVs. Despite of this the CRs are resolvable from the CVs chemically. For example, the CRs are poorer in CAIs and richer in metal than the CVs. We also remark that the sample could be strongly biased as Renazzo likely lost its most fragile presolar grains prior to accretion (Huss et al., 2003; Huss, 2004). Unfortunately, those CRs experiencing the least parent body aqueous alteration have been heavily weathered on Earth. As presolar silicates are susceptible to terrestrial weathering, future studies of CR chondrite falls would be essential in this regard.

Members of this group would be good target for searching presolar grains. In fact, Floss and Stadermann (2007) found a high abundance of presolar grains in the CR chondrite QUE99177, and inferred that this meteorite should have avoided aqueous alteration. Leitner et al. (2008) also found important amounts of presolar silicates in NWA852, and Floss and Stadermann (2008) in MET00426. Obviously, additional work on this class is required

### LL ordinary chondrites group

Some meteorites belonging to this group are particularly good candidates to find high amounts of presolar grains. They have probably experienced the lowest degree of preaccretionary processing (except for some CI and CM). The revised petrologic types given by Grossman and Brearley (2005) are particularly useful to search for good (pristine) candidates. Obvious candidates are: Semarkona (LL3.0), Bishunpur (LL3.15) and Krymka (LL3.2). Despite of this, the measured abundances of presolar silicates in the group are quite low, perhaps due to be formed from materials thermally processed in the nebula (Huss et al., 2003). We should also remark here the importance of matrix normalization. The presolar grains are found in the matrix, not in chondrules or CAIs so this effect must be considered in obtaining accurate abundances. In this regard, we note that CI chondrites are essentially 100% matrix, CMs are ~50-60% matrix, and CO and CV are ~30% matrix. Other groups, like e.g. CR, LL, L, H, and E, are ~15% matrix. Consequently a matrix normalization is required to obtain consistent results.

Future measurement of presolar grain abundances in the LL, and other ordinary chondrite groups, little bit explored at present, is urgently required.

DISCUSSION

As pointed out by Huss (1990) the abundance of refractory presolar grains in chondrites varies depending of the level of metamorphism. An extensive discussion of the issues playing a role in presolar grains preservation was made by Huss and Lewis (1995). Obviously, the initial abundances of presolar grains in the different chondrite groups were not identical, as we expect that these materials consolidated at different heliocentric distances where the materials experienced different thermal processing (Huss et al., 2003).

We think that the number of presolar silicate and oxide grains preserved in chondrites would be not only affected by metamorphism. Although minor, the role of aqueous alteration in promoting the isotopic exchange with the surrounding matrix should be explored by performing analyses of the abundances preserved in meteorites with different aqueous alteration degrees like these proposed for the CM group (Rubin et al., 2007). This process would be important as most presolar silicates are being identified during NanoSIMS analysis by looking at their peculiar isotopic anomalies over the (solar) ones exhibited by the surrounding materials forming the matrix.

What a role plays compaction in primitive meteorites? We should remark that compaction is promoted after impacts that are accompanied by shock heating. Collisional heating would also play a role in homogenising the chondrite matrixes, but all might depend of the amount of volatile material acting as a buffer of the impact-delivered heat. Trigo-Rodríguez et al. (2006) reported that the dark rims and other features observed around chondrules and inclusions in CM and other groups are likely associated with impact compaction. Recent evidence of aragonite crystals in the CM2 chondrite Murray is supporting this scenario (Lee and Ellen, 2008). On the other hand, comparing the results obtained w.r.t. the physical properties (particularly strength and bulk density) of aggregates created in the laboratory and those measured from primitive meteorites we have claimed that most solar system minor bodies have not preserved their primeval physical properties (Blum et al., 2006; Trigo-Rodríguez & Blum, 2009). Remote-sensing data on the bulk density and tensile strength of comets and meteoroids suggest that the exception could be materials coming from unprocessed comets (Blum et al., 2006).

The above mentioned picture has important implications because the precursor materials of carbonaceous chondrites possibly accreted higher amounts of volatiles than previously believed. Consequently, we cannot consider any group of carbonaceous chondrite (not even any combination of present meteorite types) as the building blocks of terrestrial planets because such bodies have evolved towards compacted samples (Nuth, 2008). This is consistent with our present picture because even on small asteroids collisionally-induced (shock) heating played an important role over Solar-System formation timescales. A recent laboratory investigation has shown that protoplanetary dust aggregates in the mm- to m-size range should be collisionally compacted within timescales $\leq 10^3$ years (Weidling et al., 2009). As collision velocities rather grow with the further evolution of planetesimals, due to the decreasing coupling to the nebular gas, we expect all asteroidal bodies to be collisionally compacted. This becomes obvious by the fact that most carbonaceous chondrites are breccias produced by impact processes (Williams et al., 1985; Rubin et al., 2007). As a consequence of impacts, shock waves accompanied by the thermal wave will propagate through the parent bodies, producing the collapse of pore spaces, and subsequent heating and shattering of the materials. These effects will participate in the progressive compaction, and this will promote the release of liquid water. The propagated thermal wave would have participated in destroying the structure of the primary minerals, but always depending of the thermal conductivity and presence of water in the original materials. The liquid element could have been present in the primordial aggregates because the mineral grains probably accreted ice and organics in the outer part of the nebula, or could be also incorporated bounded as hydrated minerals (see e.g. Brearley and Jones, 1998). We suspect that in general aqueous alteration occurred on relatively short timescales after the compaction of the progenitors. Looking for example at the aqueous alteration experienced by the CM chondrite group, the water solution that was soaking the precursors formed secondary minerals by precipitating preferentially in the pores of the matrix (Trigo-Rodríguez et al., 2006). We note that the aqueous alteration petrologic degrees suggested by Rubin et al. (2007) would be very useful to compare how aqueous alteration processes affected the survival of presolar silicates in CM and CM-like chondrites. Although the present data sample is very limited, we would ask ourselves if the abundance trend shown in Fig. 3 is associated with parent body processing. The full picture is probably more complex, but we envision that the shock heating induced by impact compaction, together with the subsequent aqueous alteration should have played an important role in survival of presolar silicates.

The effect of aqueous alteration on tiny presolar silicate grains is also consistent with the evidence that presolar silicates are much more abundant in anhydrous IDPs than in primitive meteorites. In fact, presolar silicates were first identified in IDPs up to a maximum abundance of ~5500 ppm (Messenger et al., 2003). Even higher abundances were found recently in IDPs presumably coming from comet 26P/ Grigg-Skjellerup (Nguyen et al., 2007). One should note that the parent bodies of anhydrous IDPs are probably non-thermally processed comets, such as 81P/Wild 2. Evidence for this statement has been found in the absence of hydrated minerals in recovered particles by the Stardust Preliminary Examination Team (see e.g. Brownlee et al., 2006). Despite of this, inferred abundance estimations of about 17 ppm (Stadermann and Floss, 2007) are highly biased due to the particular sampling and heating processes experienced during the capture of 81P/Wild 2 materials in Stardust aerogels. Comets are stored in the outer regions of the Solar System and can keep their starting materials intact if at their locations they never experience temperatures above the fusion-point of water ice. This condition is probably satisfied by Oort Cloud and Kuiper Belt Comets, and perhaps by some Centaurs. Future targets of missions devoted to the study of pristine bodies, like this currently under study by the European Space Agency (Marco Polo), would have particular relevance to test the abundance of presolar grains in recovered materials from a C- or D-type asteroid.

CONCLUSIONS

We have compiled the presolar grain abundances measured in primitive chondrites, making special emphasis in the recently identified presolar silicates. The following conclusions can be extracted from this work:

a) As presolar silicates are very sensitive to parent body processes (e.g. radiogenic and shock-induced metamorphism, and aqueous alteration) the measured abundances in meteorites seems to be dependent on the degree of processing experienced by the progenitors of the samples. This result is complementary to the work performed previously by Huss and Lewis (1995) and Huss et al. (2003) on the survival of interstellar diamonds and SiC.
b) Unequilibrated meteorites coming from parent bodies that experienced no aqueous alteration show the highest abundance of presolar silicates. At the top of the present list of known meteorites is the CM-related Acfer 094. Other unequilibrated COs (e.g. ALHA77307 and Y81025) are also rich in presolar silicates suggesting that those groups that escaped substantial aqueous alteration are firm candidates to preserve high abundances of presolar grains.
c) The discovery that anhydrous IDPs are probably the extraterrestrial materials containing the highest abundances of presolar silicates (Messenger et al., 2003) is perfectly consistent with our collisional-compaction scenario. Some cometary bodies formed and stored in the outer part of the nebula probably escaped the compaction experienced in the main belt. The parent bodies of primitive chondrites formed in regions subjected to higher collision rates suffered significant compaction that lead to promote chemical homogenization of matrix materials by subsequent processes (aqueous alteration and thermal metamorphism). Consequently, the abundance of presolar silicates preserved in collisionally-evolved bodies should be minor as is suggested by the currently available data. In any case, additional studies of the abundances of presolar silicates of different chondrite groups are required.
d) The observed differences in the measured presolar grains abundances in IDPs and the most unequilibrated chondrites can be explained by an inhomogeneous distribution of stellar grains in the nebula, at least among the regions of formation of chondrites and comets.
e) Assuming that the distribution of presolar grains was (opposite to d) homogeneous, the fact that highly unequilibrated chondritic meteorites have one or two orders of magnitude less abundance of presolar grains compared with IDPs, suggests a parent-body process, like e.g. the effect of shock heating and compaction of the highly-porous progenitors.

Acknowledgments

We thank an anonymous referee for a highly constructive review of the original manuscript. Our gratitude to Dr. Alan E. Rubin (UCLA) for his careful reading of a previous manuscript, and his valuable feedback to improve the manuscript's quality. J.M.T.-R also thanks *Consejo Superior de Investigaciones Científicas (CSIC)* for a JAE-Doc contract, and funding received from *Programa Nacional de Astronomía y Astrofísica* research project # AYA2008-01839/ESP.

# TABLES

Table 1. Abundance of presolar silicates in primitive chondrites, IDPs, and comet 81P/Wild 2. Temptative estimates of petrologic type are given between parentheses. Included are the statistically significant abundances found in the literature based in extensive ion microprobe rastering searches.

| Meteorite name | Group | Petrologic type | Abundance (ppm) | *References* |
|---|---|---|---|---|
| Acfer 094 | CM-like (ungrouped) | 3.0 | 180 | *Nguyen (2005); Vollmer et al. (2007)* |
| ALHA77307 | CO | 3.0 | 140 | *Nguyen (2005)* |
| MET00426 | CR |  | 120 | *Floss and Stadermann (2008)* |
| NWA852 | CR | 2 | ~90 | *Leitner et al. (2008)* |
| Y81025 | CO | 3.0 | 58-36 | *Kobayashi et al. (2005) Marhas et al. (2006)* |
| Adelaide | CV-CO | - | 41 | *Kobayashi et al. (2005)* |
| Semarkona | LL | 3.0 | 15 | *Mostefaoui et al. (2004)* |
| Bishunpur | LL | 3.1 | 15 | *Mostefaoui et al. (2004)* |
| NWA530 | CR | 2 | 3 | *Nagashima et al. (2004, 2005)* |
| Murchison | CM | 2.5* | 3 | *Nagashima et al. (2004, 2005)* |
| Tagish Lake | CI/CM (ungrouped) | 2 | 2 | *Marhas & Hoppe (2005)* |
| Mighei | CM | (2.3) | ~1** | *Zinner et al. (2003)* |
| IDPs | - | - | 375-5,500 | Messenger et al. (2003); Nguyen et al. (2007) |

*The CM petrologic subtype that takes into account aqueous alteration effects is denominated after Rubin et al. (2007). ** Presolar spinel.

# FIGURES

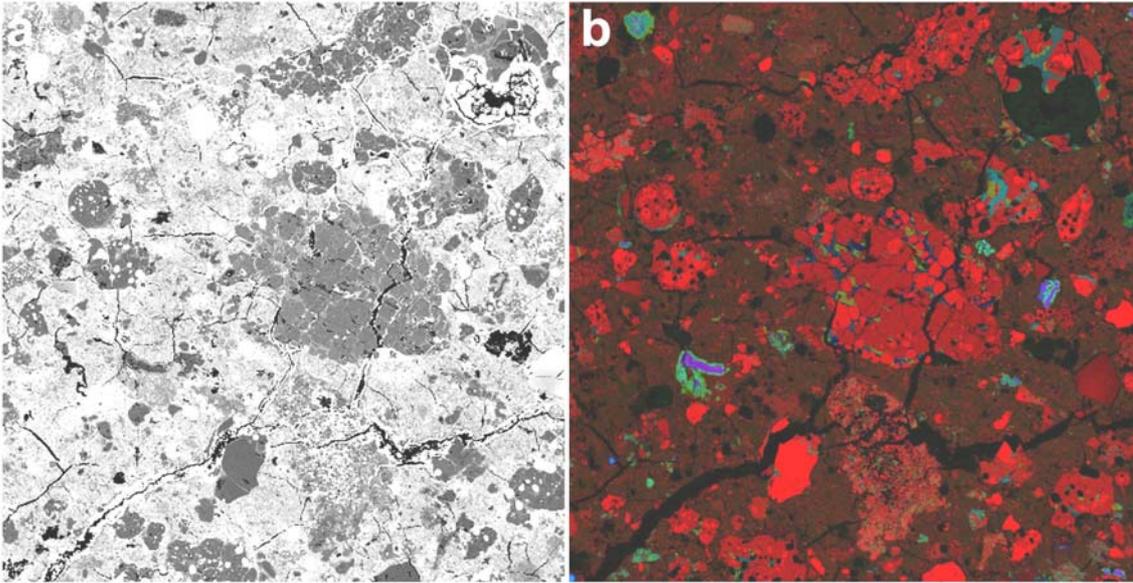

Figure 1.- A mm$^2$ area of the primitive chondrite Acfer 094. a) SEM image showing the fine texture of the matrix. b) Composite RGB ion microprobe image where red=Mg, green=Ca and blue=Al. Presolar silicate grains are typically of the size of the fine dust that is forming the matrix. The cracks that are crossing some features were produced during thin section preparation.

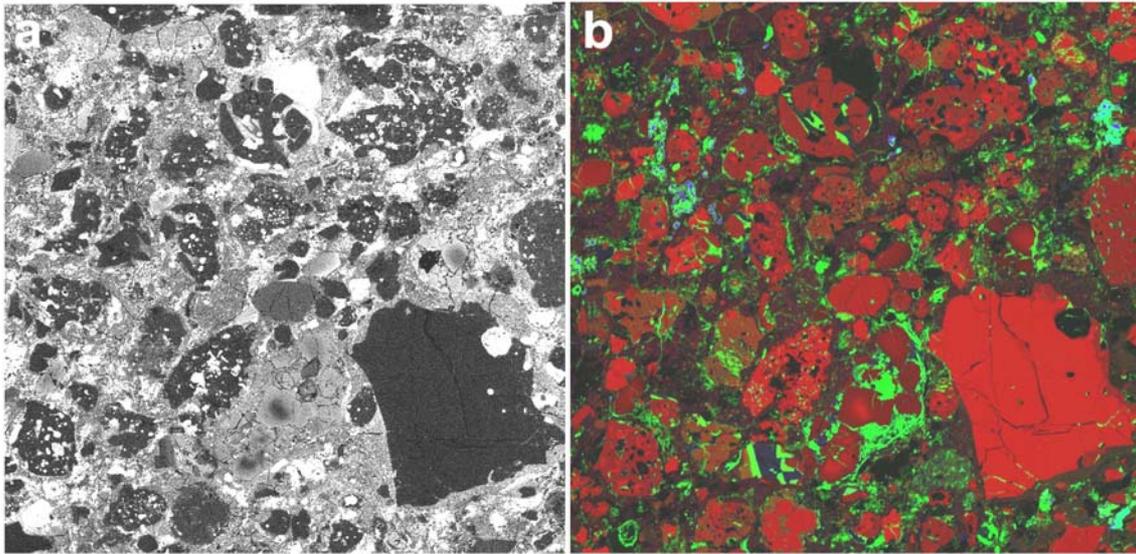

Figure 2. A mm$^2$ area of the primitive CO chondrite ALH77307. a) Backscatter electron image showing the texture of the matrix. b) Composite RGB ion microprobe image where red=Mg, green=Ca and blue=Al.

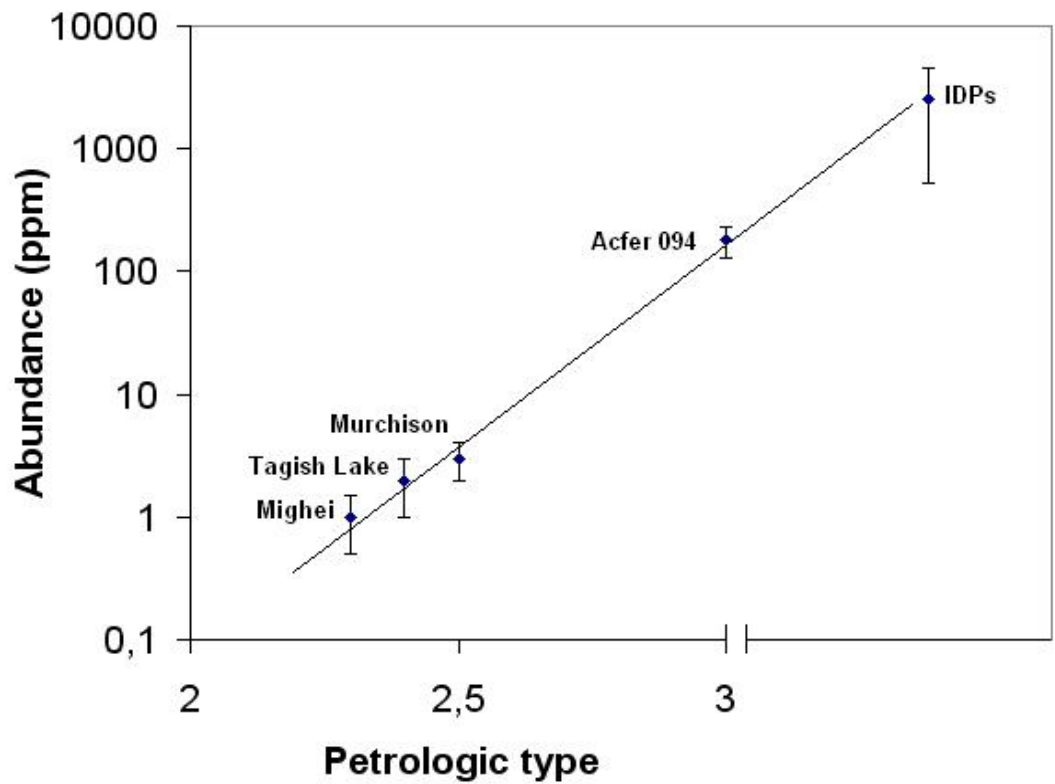

Figure 3. Measured abundances in logarithmic scale of presolar silicates in CM and CM-like chondrites versus the aqueous alteration petrologic type. Data are from Table 1. The measured abundances in IDPs are out of scale and aligned with the fit only for comparison purposes.